\documentclass[aps,prl,nofootinbib,twocolumn,superscriptaddress,preprintnumbers,floatfix]{revtex4-2} 
\usepackage[nodisplayskipstretch]{setspace}
\usepackage{cancel}
\usepackage{mathtools}

\usepackage{amssymb}
\usepackage{amsmath}
\usepackage{epsfig}
\usepackage{hyperref}
\usepackage{breakurl}
\usepackage{bm}
\usepackage{color}
\usepackage{slashed}
\usepackage{tikz}
\usetikzlibrary{decorations.markings}
\usepackage[english]{babel}
\DeclareMathOperator{\arccosh}{arccosh}

\newcommand\beq{\begin{equation}}
\newcommand\eeq{\end{equation}}
\def\bea{\begin{eqnarray}}
\def\eea{\end{eqnarray}}

\DeclareRobustCommand{\SkipTocEntry}[4]{}

\DeclareMathOperator{\arcsinh}{arcsinh}
\newcommand{\nn}{\nonumber}

\newcommand\beal{\begin{aligned}}
\newcommand\eeal{\end{aligned}}

\newcommand\dd{{\mathrm d}}
\usepackage{xcolor}

\newcommand{\bel}{{\boldsymbol \ell}}

\newcommand{\bp}{{\boldsymbol p}}
\newcommand{\bP}{{\boldsymbol P}}

\newcommand{\cD}{\mathcal{D}}

\newcommand\cS{\mathcal{S}}

\newcommand{\cR}{\mathbb {R}}


\begin{document}

\title{Radiation Reaction and Gravitational Waves at Fourth Post-Minkowskian Order} 
\author{Christoph Dlapa}
\affiliation{ Deutsches Elektronen-Synchrotron DESY, Notkestr. 85, 22607 Hamburg, Germany}

\author{Gregor K\"alin}
\affiliation{ Deutsches Elektronen-Synchrotron DESY, Notkestr. 85, 22607 Hamburg, Germany}

\author{Zhengwen Liu}
\affiliation{Niels Bohr International Academy, Niels Bohr Institute, University of Copenhagen, Blegdamsvej 17, DK-2100 Copenhagen, Denmark}
\affiliation{ Deutsches Elektronen-Synchrotron DESY, Notkestr. 85, 22607 Hamburg, Germany}

\author{Jakob Neef}
\affiliation {School of Mathematics and Statistics, University College Dublin, Belfield, Dublin 4, Ireland, D04 V1W8}
\affiliation{Humboldt-Universit\"at zu Berlin, Zum Grossen Windkanal 2, D-12489 Berlin, Germany}

\author{Rafael A. Porto}
\affiliation{ Deutsches Elektronen-Synchrotron DESY, Notkestr. 85, 22607 Hamburg, Germany}

\begin{abstract}
We obtain the total impulse in the scattering of non-spinning binaries in general relativity at fourth Post-Minkowskian order, i.e. ${\cal O}(G^4)$, including linear, nonlinear, and hereditary radiation-reaction effects. We~derive the total radiated spacetime momentum as well as the associated energy~flux. The latter can be used to compute gravitational-wave observables for generic (un)bound orbits. 
We~employ the (`in-in') Schwinger-Keldysh worldline effective field theory framework in combination with modern `multi-loop' integration techniques from collider physics. The~complete~results are in agreement with various partial calculations in the Post-Newtonian/Minkowskian expansion.\end{abstract}
 
\maketitle

{\bf Introduction.}  Waveform models are an essential ingredient in data analysis and characterization of gravitational wave (GW) signals from compact binaries~\cite{LIGOScientific:2021djp}. The~level of accuracy plays a critical role, in particular for future detectors such as LISA~\cite{LISA} and ET~\cite{ET}. In order to benefit the most from the anticipated observational reach~\cite{LISA,ET,buosathya,tune,music,Maggiore:2019uih,Barausse:2020rsu,Bernitt:2022aoa}, the modelling of GW sources must therefore continue to develop---both through analytic methodologies~\cite{Damour:2008yg,blanchet,Schafer:2018kuf,walterLH,iragrg,review,Barack:2018yvs,Buonanno:2022pgc,Goldberger:2022ebt} and numerical simulations~\cite{Ajith:2012az,Szilagyi:2015rwa,Dietrich:2018phi}---in parallel with the expected increase in sensitivity with next-generation GW interferometers.\vskip 4pt Motivated~by the {\it Effective-One-Body} (EOB) formalism~\cite{Bini:2012ji,damour1,damour2,Damour:2019lcq,Damour:2020tta,Bini:2021gat}, the {\it Boundary-to-Bound} (B2B) dictionary between unbound and bound observables \cite{paper1,paper2,b2b3}, and benefiting from powerful `multi-loop' integrations tools~\cite{Smirnov:2012gma,Kotikov:1991pm,Remiddi:1997ny,Henn:2013pwa,Prausa:2017ltv,Lee:2020zfb,Lee:2014ioa,Adams:2018yfj,Chetyrkin:1981qh,Tkachov:1981wb,Smirnov:2019qkx,Smirnov:2020quc,Lee:2012cn,Lee:2013mka,Beneke:1997zp,Jantzen:2012mw,Smirnov:2015mct,Meyer:2016zeb,Meyer:2016slj,Broedel:2019kmn,Primo:2017ipr,Hidding:2020ytt,Goncharov:2001iea,Chen:1977oja,Duhr:2014woa,Duhr:2019tlz,Dlapa:2020cwj,Smirnov:2021rhf,Lee:2019zop,Blumlein:2021pgo,dklp}, significant progress has been achieved in recent years in our analytic understanding of (classical) gravitational scattering in the Post-Minkowskian (PM) expansion in powers of $G$ (Newton's constant); both via effective field theory (EFT)~\cite{pmeft,3pmeft,tidaleft,pmefts,4pmeft,4pmeft2,janmogul,janmogul2,Jakobsen:2022fcj,Jakobsen:2021zvh,Mougiakakos:2021ckm,Riva:2021vnj,Mougiakakos:2022sic,Riva:2022fru,eftrad,Jakobsen:2022psy,Jinno:2022sbr} and amplitude-based~\cite{Amati:1990xe,ira1,Vaidya:2014kza,Goldberger:2016iau,cheung,Ciafaloni:2018uwe,Guevara:2018wpp,donal,zvi1,Haddad:2020que,Brandhuber:2021eyq,Aoude:2022thd,Bjerrum-Bohr:2021din,4pmzvi,4pmzvi2,Gabriele2,parra,FebresCordero:2022jts,Parra3,Manohar:2022dea} methodologies. The PM regime incorporates an infinite tower of Post-Newtonian (PN) corrections at a given order in $G$ that may increase the accuracy of phenomenological waveform models~\cite{Khalil:2022ylj,Hopper:2022rwo}.\vskip 4pt However, despite some notable exceptions  \cite{Kovacs:1978eu,Amati:1990xe,Ciafaloni:2018uwe,Parra3,Damour:2020tta,Riva:2021vnj,Mougiakakos:2022sic,Riva:2022fru,Manohar:2022dea,Jakobsen:2022psy,eftrad}, the majority of the PM computations have so far impacted our knowledge of the {\it conservative} sector, with potential interactions \cite{4pmeft,4pmzvi} as well as `tail effects'~\cite{4pmeft2,4pmzvi2} known to 4PM order.~Yet, until now, complete results had not been obtained at the same level of accuracy. The purpose of this letter is therefore to~report the total change of (mechanical) momentum, a.k.a.~the impulse, for the gravitational scattering of non-spinning bodies---including all the hitherto unknown linear, nonlinear and hereditary radiation-reaction {\it dissipative} effects---at ${\cal O}(G^4)$, from which we derive the total radiated spacetime momentum and GW energy flux.\vskip 4pt Building on pioneering developments in the PN~regime \cite{nrgr,chadprl,chadRR,Galley:2010es,chadbr2,natalia1,natalia2,tail}, the derivation proceeds via the EFT approach in a PM scheme \cite{pmeft}, extended~in~\cite{eftrad}~to simultaneously incorporate conservative and dissipative effects via the `in-in' Schwinger-Keldysh formalism~\cite{Schwinger:1960qe,Keldysh:1964ud,Calzetta:1986ey,Calzetta:1986cq,Jordan:1986ug}. As~discussed in \cite{eftrad}, the in-in~impulse resembles the `in-out' counterpart used in the conservative sector \cite{4pmeft,4pmeft2}, except for its causal structure which entails the use of retarded Green's functions \cite{eftrad}. After adapting integration tools to our problem, the calculation of the impulse is mapped to a series of `three-loop'  mass-independent integrals. As~in previous derivations \cite{pmeft,3pmeft,4pmeft,4pmeft2}, the latter are solved via the methodology of differential equations~\cite{Kotikov:1991pm,Remiddi:1997ny,Henn:2013pwa,Prausa:2017ltv,Lee:2020zfb,Lee:2014ioa,Adams:2018yfj}. The~relativistic two-body problem is then reduced to obtaining the necessary boundary conditions in the near-static limit. The boundary integrals are computed using the method of regions \cite{Beneke:1997zp}, involving potential (off-shell) and radiation (on-shell) modes \cite{nrgr}. The full solution is thus {\it bootstrapped} to all orders in the velocities from the same type of calculations needed in the EFT approach with PN sources~\cite{nrgr,tail,nrgr4pn1,nrgr4pn2,Blumlein:2021txe,Almeida:2022jrv}. As~a nontrivial check, by rewriting retarded Green's functions as Feynman propagators plus a {\it reactive} term \cite{eftrad}, we recover the value in \cite{4pmeft,4pmeft2} for the Feynman-only (conservative) part. Agreement~is also found in the overlap with various PN derivations \cite{Damour:2020tta,Cho:1,Cho:2,Bini:2021gat,Bini:2021qvf,Bini:2022yrk,Bini:2022xpp,privateDD} and partial PM results \cite{Manohar:2022dea} obtained using the relations in~\cite{Bini:2012ji}. 
\vskip 4pt

The B2B dictionary \cite{paper1,paper2,b2b3} allows us to connect scattering data to observables for bound states via analytic continuation.  However, similarly to the lack of periastron advance at 3PM \cite{paper1,paper2,3pmeft}, the symmetries of the problem yield a vanishing coefficient for the radiated energy integrated over a period of elliptic-like motion at 4PM, trivially recovered by the B2B map. Nevertheless, since nonlinear radiation-reaction effects do not contribute to the integrated radiated energy at ${\cal O}(G^4)$, we can then derive the GW flux in an adiabatic approximation~\cite{b2b3}. This allows for the computation of radiative observables for generic (un)bound orbits through balance equations,~as in the EOB approach \cite{Bini:2012ji}, thus including an infinite series of velocity corrections.\vskip 4pt

\begin{figure}[t!]
  \includegraphics[width=0.7\linewidth]{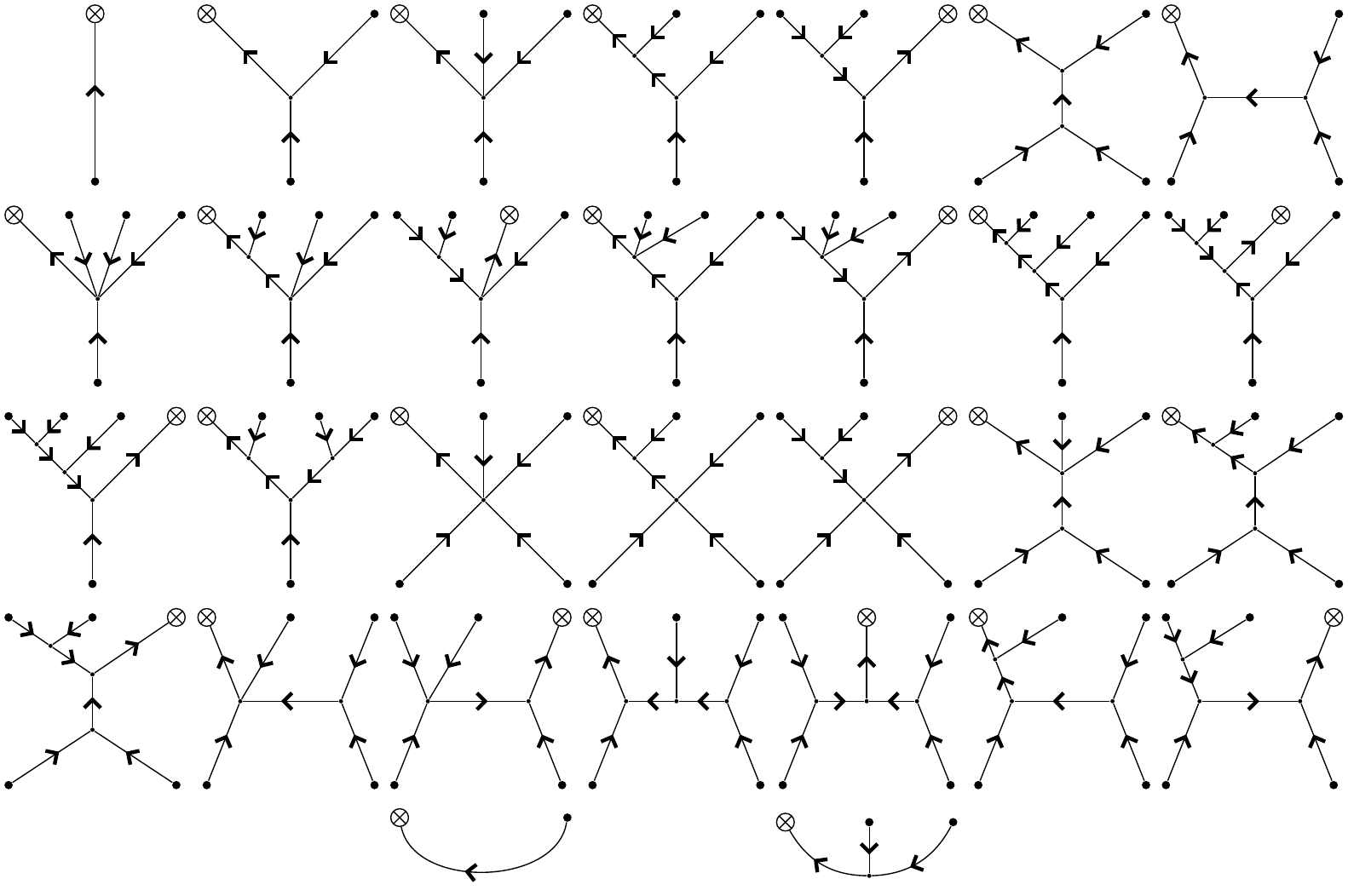}
  \vskip -1em
  \caption{In-in Feynman topologies to 4PM order. The arrows indicate the flow of (retarded) time. The crosses represent the location of the derivative in the impulse in \eqref{impulse0}. The last two diagrams are the only `self-energies' needed at 4PM \cite{eftrad}.}
  \vspace{-0.2cm}
  \label{fig1}
\end{figure}

 {\bf The EFT} {\it in-in}{\bf tegrand.} Following the Schwinger-Keldysh formalism \cite{Schwinger:1960qe,Keldysh:1964ud,Calzetta:1986ey,Calzetta:1986cq,Jordan:1986ug} adapted to the EFT approach in \cite{chadRR,eftrad}, the effective action is obtained via a {\it closed-time-path} integral involving a doubling of the metric perturbation $(h_{\mu\nu}^\pm)$ as well as the worldline ($x^\alpha_{a,\pm}$) degrees of freedom, schematically, 
 \beq
 \label{seff}
e^{i\cS_{\rm eff}[x_{a,\pm}]} = \int \cD h^+ \cD h^- \, e^{i \left\{S_{\rm EH}[h^\pm] + S_{\rm pp}[h^\pm,x_{a,\pm}]\right\}} \,,
 \eeq
 with $S_{\rm EH}$ and $S_{\rm pp}$ the closed-path version of the Einstein-Hilbert and point-particle worldline actions, respectively.  We ignore here spin degrees of freedom and finite-size effects (see \cite{pmeft,pmefts}). We also restrict ourselves to the classical regime and therefore the path integral in \eqref{seff} is computed in the saddle-point approximation---keeping only connected `tree-level' Feynman diagrams of the gravitational field(s)---with the compact objects treated as external nonpropagating sources. \vskip 4pt  In this scenario, the matrix of (causal) propagators is given by the Keldysh representation:
   \begin{equation}
  K^{AB}(x-y) = i \begin{pmatrix} 0 & -\Delta_{\textnormal{adv}}(x-y) \\ -\Delta_{\textnormal{ret}}(x-y) & 0 \end{pmatrix}\,,\label{kmatrix}
\end{equation}
with $A,B \in\{+,-\}$ and $\Delta_{\rm ret/adv}$ the standard retarded/advanced Green's functions. The impulse, e.g. for particle 1, then follows from \cite{eftrad} 
 \beq
  \label{impulse0}
    \Delta p_1^\mu \!=\!
    \left.-\eta^{\mu\nu} \!\int_{-\infty}^\infty\! \dd\tau_1 \frac{\delta\cS_\textrm{eff}[x_{a,\pm}]}{\delta x_{1,-}^\nu(\tau_1)}\right|_{\rm PL}\!\! =\!
    \sum_n  G^n \Delta^{(n)}p_1^\mu\,,
\eeq  
to all PM orders, where the subscript `PL' stands for the {\it Physical Limit}: $\{x_{a,-}\to~0,\, x_{a,+}\to x_a\}$ \cite{chadprl}. As for the conservative sector \cite{4pmeft,4pmeft2}, we~must also include {\it iterations} from lower order solutions to the equations of motion. The latter are obtained from the effective action in the same physical limit. The diagrams needed to ${\cal O}(G^4)$ are depicted in Fig.~\ref{fig1}. Mirror images (not shown) must also be computed. See \cite{eftrad} for details.\vskip 4pt 
 The impulse is further decomposed into scalar integrals in the perpendicular and longitudinal directions, i.e. for particle 1 (and likewise for particle 2)
\beq
\label{impulse}
\Delta^{(n)} p_1^\mu =c^{(n)}_{1b}\, \frac{\hat b^\mu}{b^n} + \frac{1}{b^n} \sum_{a} c^{(n)}_{1\check{u}_a}\, \check{u}_a^\mu \,,
\eeq
with $b^\mu\equiv b^\mu_1-b_2^\mu$ the impact parameter, $b \equiv \sqrt{-b^\mu b_\mu}$, and $\hat b^\mu \equiv b^\mu/b$. We use the notation~\cite{Parra3}
\beq
 \check{u}^\mu_1 \equiv \frac{\gamma u^\mu_2-u^\mu_1}{\gamma^2-1}\,, \quad  \check{u}^\mu_2 \equiv \frac{\gamma u^\mu_1 - u^\mu_2}{\gamma^2-1}\,,\quad \gamma \equiv u_1\cdot u_2\,,
 \eeq
 with $u_a$'s the incoming velocities, $b\cdot u_a\!=\!0$, $u_a^2\!=\!1$, and $\check{u}_a\cdot u_b = \delta_{ab}$. Ignoring absorption, the preservation of the on-shell condition, $p_a^2=m_a^2$,  implies $2p_a\cdot \Delta p_a  = -  (\Delta p_a)^2$, which serves as a nontrivial consistency check.\vskip 4pt

 {\bf Integration.}  Similarly to the derivations in \cite{4pmeft,4pmeft2}, but incorporating the key distinction between Feynman and retarded propagators, the components of the impulse can be reduced to different families of integrals,
\begin{align}\label{4pm-ints}
\int \prod_{i=1}^3  \frac{\mathrm{d}^d \ell_i}{\pi^{d/2}}
\frac{\delta(\ell_i \!\cdot\! u_{a_i})}  { 
(\pm \ell_i \!\cdot\! u_{\slashed{a}_i} {-} i0)^{n_i}}\prod_{k=1}^{9} {1 \over  D_k^{\nu_k}}\,,
\end{align}
restricted by  Dirac-$\delta$ functions. Following~\cite{4pmeft,4pmeft2}, the $\ell_{i=1,2,3}$'s are the loop momenta, $n_i,\nu_k$ are integers, and $a_i\in\{1,2\}$, with $u_\slashed{1}=u_2$, $u_\slashed{2}=u_1$. In contrast to the conservative part, the $D_k$'s are now various sets of retarded/advanced propagators, e.g. $\{(\ell^0\pm i0)^2-\bel^2,\ldots\}$, consistent with causality. The~same constraints as before apply on the external data \cite{3pmeft}, such that the relevant integrals can only depend on~$\gamma$. As~in our previous calculations \cite{4pmeft,4pmeft2}, we conveniently introduce the parameter~$x$, defined through the relation $\gamma \equiv (x^2 {+} 1)/2x$~\cite{parra}, and compute these integrals by using dimensional regularization (in $d \equiv 4-2\epsilon$ dimensions) and the method of differential equations \cite{Kotikov:1991pm,Remiddi:1997ny,Henn:2013pwa,Lee:2014ioa, Prausa:2017ltv, Adams:2018yfj, Lee:2020zfb}.\vskip 4pt  The integration problem then resembles the steps already performed for the computation in \cite{4pmeft,4pmeft2}, except for a few notable differences. First of all, as before we use integration-by-parts (IBP) relations \cite{Chetyrkin:1981qh, Tkachov:1981wb,Smirnov:2019qkx,Smirnov:2020quc,Lee:2012cn,Lee:2013mka} and reduce \eqref{4pm-ints} to a basis of master integrals. Because of the fewer number of symmetries of the in-in integrand, the algebraic manipulations become a bit more involved than with Feynman-only propagators. But more importantly,  the boundary conditions in the near-static limit $\gamma \simeq 1$ must be computed in terms of retarded/advanced Green's functions. For this purpose, we resort to the method of regions and the expansion into potential and radiation modes. The potential-only part was obtained in \cite{4pmeft}, and recovered here from the full solution. For radiation modes, the same type of integrals appearing in PN derivations \cite{tail}, combined with leftover integrals over potential-only modes at one and two loops, are sufficient to bootstrap the entire answer.~See~\cite{dklp} for more details.

\begin{widetext}

{\bf Total impulse.} Inputing the values of the boundary master integrals and translating from $x$ to $\gamma$ space, we find
  \beq\label{imptot}
  \resizebox{0.95\hsize}{!}{
    $
    \begin{aligned}
      \frac{c^{(4)\rm tot}_{1b}}{\pi} & \begin{multlined}[t]=
        -\frac{3 h_1 m_1 m_2 (m_1^3+m_2^3)}{64 (\gamma ^2-1)^{5/2}}
        +m_1^2 m_2^2 (m_1+m_2)\bigg[
          \frac{21 h_2 \mathrm{E}^2\left(\frac{\gamma -1}{\gamma +1}\right)}{32 (\gamma -1) \sqrt{\gamma ^2-1}}
          +\frac{3 h_3 \mathrm{K}^2\left(\frac{\gamma -1}{\gamma +1}\right)}{16 \left(\gamma ^2-1\right)^{3/2}}
          -\frac{3 h_4 \mathrm{E}\left(\frac{\gamma -1}{\gamma +1}\right) \mathrm{K}\left(\frac{\gamma -1}{\gamma +1}\right)}{16 \left(\gamma ^2-1\right)^{3/2}}
          +\frac{\pi ^2 h_5}{8 \sqrt{\gamma ^2-1}}
          +\frac{h_6 \log \left(\frac{\gamma -1}{2}\right)}{16 \left(\gamma ^2-1\right)^{3/2}}\\
          +\frac{3 h_7 \text{Li}_2\left(\sqrt{\frac{\gamma -1}{\gamma +1}}\right)}{(\gamma -1) (\gamma +1)^2}
          -\frac{3 h_7 \text{Li}_2\left(\frac{\gamma -1}{\gamma +1}\right)}{4 (\gamma -1) (\gamma +1)^2}
          \bigg]
        +m_1^3 m_2^2\bigg[
          \frac{h_8}{48 \left(\gamma ^2-1\right)^3}
          +\frac{\sqrt{\gamma ^2-1} h_9}{768 (\gamma -1)^3 \gamma ^9 (\gamma +1)^4}
          +\frac{h_{10} \log \left(\frac{\gamma +1}{2}\right)}{8 \left(\gamma ^2-1\right)^2}
          -\frac{h_{11} \log \left(\frac{\gamma +1}{2}\right)}{32 \left(\gamma ^2-1\right)^{5/2}}
          +\frac{h_{12} \log (\gamma )}{16 \left(\gamma ^2-1\right)^{5/2}}\\
          -\frac{h_{13}\arccosh(\gamma )}{8 (\gamma -1) (\gamma +1)^4}
          +\frac{h_{14}\arccosh(\gamma )}{16 \left(\gamma ^2-1\right)^{7/2}}
          -\frac{3 h_{15} \log \left(\frac{\gamma +1}{2}\right) \log \left(\frac{\gamma -1}{\gamma +1}\right)}{8 \sqrt{\gamma ^2-1}}
          +\frac{3 h_{16}\arccosh(\gamma ) \log \left(\frac{\gamma -1}{\gamma +1}\right)}{16 \left(\gamma ^2-1\right)^2}
          -\frac{3 h_{17} \text{Li}_2\left(\frac{\gamma -1}{\gamma +1}\right)}{64 \sqrt{\gamma ^2-1}}
          -\frac{3}{32} \sqrt{\gamma ^2-1} h_{18} \text{Li}_2\left(\frac{1-\gamma }{\gamma +1}\right)
          \bigg]\\
        +m_1^2 m_2^3 \bigg[
          \frac{3 h_{15} \log \left(\frac{2}{\gamma -1}\right) \log \left(\frac{\gamma +1}{2}\right)}{8 \sqrt{\gamma ^2-1}}
          +\frac{3 h_{16} \log \left(\frac{\gamma -1}{2}\right)\arccosh(\gamma )}{16 \left(\gamma ^2-1\right)^2}
          +\frac{h_{19}}{48 \left(\gamma ^2-1\right)^3}
          +\frac{h_{20}}{192 \gamma ^7 \left(\gamma ^2-1\right)^{5/2}}
          +\frac{h_{21} \log \left(\frac{\gamma +1}{2}\right)}{8 \left(\gamma ^2-1\right)^2}
          +\frac{h_{22} \log \left(\frac{\gamma +1}{2}\right)}{16 \left(\gamma ^2-1\right)^{3/2}}
          +\frac{h_{23} \log (\gamma )}{2 \left(\gamma ^2-1\right)^{3/2}}\\
          -\frac{h_{24}\arccosh(\gamma )}{16 \left(\gamma ^2-1\right)^3}
          +\frac{h_{25}\arccosh(\gamma )}{16 \left(\gamma ^2-1\right)^{7/2}}
          -\frac{3 h_{26}\arccosh^2(\gamma )}{32 \left(\gamma ^2-1\right)^{7/2}}
          +\frac{3 h_{27} \log ^2\left(\frac{\gamma +1}{2}\right)}{2 \sqrt{\gamma ^2-1}}
          +\frac{3 h_{28} \log \left(\frac{\gamma +1}{2}\right)\arccosh(\gamma )}{16 \left(\gamma ^2-1\right)^2}
          +\frac{h_{29} \text{Li}_2\left(\frac{1-\gamma }{\gamma +1}\right)}{4 \sqrt{\gamma ^2-1}}
          +\frac{3 h_{30} \text{Li}_2\left(\frac{\gamma -1}{\gamma +1}\right)}{8 \sqrt{\gamma ^2-1}}
          \bigg]\,,
      \end{multlined}\\
      c^{(4)\rm tot}_{1\check{u}_1} &=
      \frac{9 \pi ^2 h_{31} m_1 m_2^2 \left(m_1+m_2\right){}^2}{32 \left(\gamma ^2-1\right)}
      +\frac{2 h_{32} m_1 m_2^2 \left(m_1^2+m_2^2\right)}{\left(\gamma ^2-1\right)^3}
      +m_1^2 m_2^3\bigg[
        \frac{4 h_{33}}{3 \left(\gamma ^2-1\right)^3}
        -\frac{8 h_{34}}{3 \left(\gamma ^2-1\right)^{5/2}}
        +\frac{8 h_{35}\arccosh(\gamma )}{\left(\gamma ^2-1\right)^3}
        -\frac{16 h_{36}\arccosh(\gamma )}{\left(\gamma ^2-1\right)^{3/2}}
        \bigg]\,,\\
      c^{(4)\rm tot}_{1\check{u}_2} &\begin{multlined}[t]=
      -m_1^4 m_2 \left(\frac{9 \pi ^2 h_{31}}{32 \left(\gamma ^2-1\right)}+\frac{2 h_{32}}{\left(\gamma ^2-1\right)^3}\right)
      +m_1^3 m_2^2\bigg[
        -\frac{4 h_{37}}{3 \left(\gamma ^2-1\right)^3}
        +\frac{h_{38}}{705600 \gamma ^8 \left(\gamma ^2-1\right)^{5/2}}
        +\frac{\pi ^2 h_{39}}{192 \left(\gamma ^2-1\right)^2}
        +\frac{h_{40}\arccosh(\gamma )}{6720 \gamma ^9 \left(\gamma ^2-1\right)^3}
        +\frac{32 h_{41}\arccosh(\gamma )}{3 \left(\gamma ^2-1\right)^{3/2}}\\
        -\frac{8 h_{42}\arccosh^2(\gamma )}{\left(\gamma ^2-1\right)^2}
        +\frac{32 h_{43}\arccosh^2(\gamma )}{\left(\gamma ^2-1\right)^{7/2}}
        +\frac{h_{44} \log (2)\arccosh(\gamma )}{8 \left(\gamma ^2-1\right)^2}
        +\frac{3 h_{45} \left(\text{Li}_2\left(\frac{\gamma -1}{\gamma +1}\right)-4 \text{Li}_2\left(\sqrt{\frac{\gamma -1}{\gamma +1}}\right)\right)}{16 \left(\gamma ^2-1\right)^2}\\
        +\frac{3 h_{46} \left(\log \left(\frac{\gamma +1}{2}\right)\arccosh(\gamma )-2 \text{Li}_2\left(\sqrt{\gamma ^2-1}-\gamma \right)\right)}{8 \left(\gamma ^2-1\right)^2}
        -\frac{h_{47} \left(\text{Li}_2\left(-\left(\gamma -\sqrt{\gamma ^2-1}\right)^2\right)-2 \log (\gamma )\arccosh(\gamma )\right)}{16 \left(\gamma ^2-1\right)^2}
        \bigg]
      +m_1^2 m_2^3\bigg[
        -\frac{2 h_{48}}{45 \left(\gamma ^2-1\right)^3}\\
        +\frac{h_{49}}{1440 \gamma ^7 \left(\gamma ^2-1\right)^{5/2}}
        +\frac{\pi ^2 h_{50}}{48 \left(\gamma ^2-1\right)^2}
        +\frac{h_{51}\arccosh(\gamma )}{480 \gamma ^8 \left(\gamma ^2-1\right)^3}
        -\frac{16 h_{52}\arccosh(\gamma )}{5 \left(\gamma ^2-1\right)^{3/2}}
        -\frac{16 h_{53}\arccosh^2(\gamma )}{\left(\gamma ^2-1\right)^2}
        -\frac{32 h_{54}\arccosh^2(\gamma )}{\left(\gamma ^2-1\right)^{7/2}}
        -\frac{h_{55} \log (2)\arccosh(\gamma )}{4 \left(\gamma ^2-1\right)^2}\\
        +\frac{h_{56} \left(\text{Li}_2\left(\frac{\gamma -1}{\gamma +1}\right)-4 \text{Li}_2\left(\sqrt{\frac{\gamma -1}{\gamma +1}}\right)\right)}{32 \left(\gamma ^2-1\right)^2}
        +\frac{h_{57} \left(\log \left(\frac{2}{\gamma +1}\right)\arccosh(\gamma )+2 \text{Li}_2\left(\sqrt{\gamma ^2-1}-\gamma \right)\right)}{4 \left(\gamma ^2-1\right)^2}
        +\frac{h_{58} \left(\text{Li}_2\left(-\left(\gamma -\sqrt{\gamma ^2-1}\right)^2\right)-2 \log (\gamma )\arccosh(\gamma )\right)}{8 \left(\gamma ^2-1\right)^2}
        \bigg]\,.
    \end{multlined}
\end{aligned}$
}
\eeq
See Table \ref{tbl:coeffs} for the list of $h_i(\gamma)$ polynomials.
\end{widetext}
The impulse for the second particle follows by exchanging $1\leftrightarrow 2$ in the masses, incoming velocities, and impact parameter. As expected from the calculations in \cite{4pmeft,4pmeft2}, the complete results feature dilogarithms~($\text{Li}_2(z)$), and complete elliptic integrals of the first $(\mathrm{K}(z))$ and second ($\mathrm{E}(z)$)~kind.

\vskip 4pt {\it Conservative.} As shown explicitly in \cite{eftrad}, a (time-symmetric) conservative contribution can be identified by rewriting retarded Green's functions in terms of Feynman propagators plus a reactive term, and keeping the real part of the Feynman-only piece, i.e. $\Delta p^\mu_{\rm cons} \equiv  \cR \Delta p^\mu_{\rm F}$, with imaginary terms canceling out against counter-parts from the reactive terms. Performing these steps in the full in-in integrand, and associated boundary conditions entering in the total impulse, we readily recover the conservative results in \cite{4pmeft,4pmeft2} including potential and radiation-reaction tail effects. 

\vskip 4pt {\it Dissipative.}  As discussed in \cite{eftrad}, the terms stemming off of the mismatch between Feynman and retarded propagators incorporate dissipative effects. Needless to say, these terms can also be read off directly from the total result by subtracting the conservative part. Following the analysis in \cite{4pmeft,4pmeft2}, we disentangle the various pieces according to factors of $v_\infty^{-2\epsilon}$, with $v_\infty \equiv \sqrt{\gamma^2-1}$, which signal the presence of an on-shell mode.\vskip 4pt 

Starting with a single radiation mode we encounter  {\it instantaneous} dissipative effects at linear order in the radiation-reaction. The latter are odd under time reversal and contribute to the $\hat b$ and $\check{u}_a$ directions. We~find agreement in the overlap with known partial results in linear-response theory in the PN literature~\cite{Damour:2020tta,Cho:1,Cho:2,Bini:2022yrk,Bini:2022xpp,Bini:2021qvf,Bini:2021gat}, as well as with the (odd) contribution to the $c^{(4)}_{1b}$ coefficient recently derived in \cite{Manohar:2022dea}. All of the remaining radiative terms involve two radiation modes. After~removing the Feynman-only (radiative) conservative pieces \cite{4pmeft,4pmeft2}, the leftovers contain hereditary as well as nonlinear radiation-reaction dissipative effects. The former enters both in the longitudinal and perpendicular directions, whereas the latter contributes only to the total radiated perpendicular momentum at this order.\footnote{At this point, however, we cannot distinguish whether nonlinear radiation-reaction terms  are due to either effects at second order in the linear radiation-reaction force or truly nonlinear gravitational corrections.} We~also find perfect consistency with known nonlinear and hereditary results in the PN expansion \cite{Cho:2,Bini:2021qvf,Bini:2022yrk,Bini:2022xpp,privateDD}.\vskip 4pt See~the supplemental material and~ancillary~file for explicit expressions.
\begin{widetext}
  \begin{table*}[tb]
    \centering
    \scalebox{0.6}{
      \begin{tabular}{|p{12.7cm}|p{12.7cm}|}
        \hline
        \vspace{-15pt}
        \begin{equation}
          \begin{aligned}
            h_1 &= 515 \gamma ^6-1017 \gamma ^4+377 \gamma ^2-3\\
            h_2 &= 380 \gamma ^2+169\\
            h_3 &= 1200 \gamma ^2+2095 \gamma +834\\
            h_4 &= 1200 \gamma ^3+2660 \gamma ^2+2929 \gamma +1183\\
            h_5 &= -25 \gamma ^6+30 \gamma ^4+60 \gamma ^3-129 \gamma ^2+76 \gamma -12\\
            h_6 &= 210 \gamma ^6-552 \gamma ^5+339 \gamma ^4-912 \gamma ^3+3148 \gamma ^2-3336 \gamma +1151\\
            h_7 &= -\gamma  \left(2 \gamma ^2-3\right) \left(15 \gamma ^2-15 \gamma +4\right)\\
            h_8 &= 420 \gamma ^9+3456 \gamma ^8-1338 \gamma ^7-15822 \gamma ^6+13176 \gamma ^5+9563 \gamma ^4-16658 \gamma ^3\\
            &\quad+8700 \gamma ^2-496 \gamma -1049\\
            h_9 &= -22680 \gamma ^{21}+11340 \gamma ^{20}+116100 \gamma ^{19}-34080 \gamma ^{18}-216185 \gamma ^{17}+74431 \gamma ^{16}\\
            &\quad+232751 \gamma ^{15}-304761 \gamma ^{14}+333545 \gamma ^{13}-32675 \gamma ^{12}-500785 \gamma ^{11}+535259 \gamma ^{10}\\
            &\quad-181493 \gamma ^9+3259 \gamma ^8+9593 \gamma ^7+9593 \gamma ^6-3457 \gamma ^5-3457 \gamma ^4\\
            &\quad+885 \gamma ^3+885 \gamma ^2-210 \gamma -210\\
            h_{10} &= -280 \gamma ^7+50 \gamma ^6+970 \gamma ^5+27 \gamma ^4-1432 \gamma ^3+444 \gamma ^2+366 \gamma -129\\
            h_{11} &= 2835 \gamma ^{11}-10065 \gamma ^9-700 \gamma ^8+13198 \gamma ^7+1818 \gamma ^6-9826 \gamma ^5+5242 \gamma ^4\\
            &\quad+11391 \gamma ^3+18958 \gamma ^2+10643 \gamma +2074\\
            h_{12} &= \gamma  \left(945 \gamma ^{10}-2955 \gamma ^8+4874 \gamma ^6-5014 \gamma ^4+8077 \gamma ^2+5369\right)\\
            h_{13} &= \gamma  \left(280 \gamma ^7+580 \gamma ^6+90 \gamma ^5-856 \gamma ^4-2211 \gamma ^3+1289 \gamma ^2+2169 \gamma -1965\right)\\
            h_{14} &= \gamma  \left(2 \gamma ^2-3\right) \left(280 \gamma ^7-890 \gamma ^6-610 \gamma ^5+1537 \gamma ^4+380 \gamma ^3-716 \gamma ^2-82 \gamma +85\right)\\
            h_{15} &= 35 \gamma ^4+60 \gamma ^3-150 \gamma ^2+76 \gamma -5\\
            h_{16} &= \gamma  \left(2 \gamma ^2-3\right) \left(35 \gamma ^4-30 \gamma ^2+11\right)\\
            h_{17} &= 315 \gamma ^8-860 \gamma ^6+690 \gamma ^4-960 \gamma ^3+1732 \gamma ^2-1216 \gamma +299\\
            h_{18} &= 315 \gamma ^6-145 \gamma ^4+65 \gamma ^2+21\\
            h_{19} &= 840 \gamma ^9+1932 \gamma ^8+234 \gamma ^7-17562 \gamma ^6+20405 \gamma ^5-2154 \gamma ^4-11744 \gamma ^3\\
            &\quad+12882 \gamma ^2-4983 \gamma +102\\
            h_{20} &= 3600 \gamma ^{16}+4320 \gamma ^{15}-23840 \gamma ^{14}+7824 \gamma ^{13}+14128 \gamma ^{12}+16138 \gamma ^{11}-9872 \gamma ^{10}\\
            &\quad-47540 \gamma ^9+63848 \gamma ^8-37478 \gamma ^7+13349 \gamma ^6-1471 \gamma ^4+207 \gamma ^2-45\\
            h_{21} &= -350 \gamma ^7+1425 \gamma ^5-400 \gamma ^4-1480 \gamma ^3+660 \gamma ^2+285 \gamma -124\\
            h_{22} &= -300 \gamma ^7+210 \gamma ^6+1112 \gamma ^5+2787 \gamma ^4+2044 \gamma ^3+3692 \gamma ^2+6744 \gamma +1759\\
            h_{23} &= \gamma  \left(75 \gamma ^6-140 \gamma ^4-283 \gamma ^2-852\right)\\
            h_{24} &= \gamma  \left(2 \gamma ^2-3\right) \left(210 \gamma ^6-720 \gamma ^5+339 \gamma ^4-576 \gamma ^3+3148 \gamma ^2-3504 \gamma +1151\right)\\
            h_{25} &= \gamma  \left(2 \gamma ^2-3\right) \left(350 \gamma ^7-960 \gamma ^6-705 \gamma ^5+1632 \gamma ^4+432 \gamma ^3-768 \gamma ^2-93 \gamma +96\right)\\
            h_{26} &= \gamma ^2 \left(3-2 \gamma ^2\right)^2 \left(35 \gamma ^4-30 \gamma ^2+11\right)\\
            h_{27} &= 15 \gamma ^3+60 \gamma ^2+19 \gamma +8\\
            h_{28} &= \gamma  \left(70 \gamma ^6-645 \gamma ^4+768 \gamma ^2+63\right)\\
            h_{29} &= -75 \gamma ^6+90 \gamma ^4+333 \gamma ^2+60
          \end{aligned}\nn
        \end{equation}
        \vspace{-11pt}
        &
        \vspace{-15pt}
        \begin{equation}
          \begin{aligned}
            h_{30} &= 25 \gamma ^6-30 \gamma ^4+60 \gamma ^3+129 \gamma ^2+76 \gamma +12\\
            h_{31} &= \left(1-5 \gamma ^2\right)^2\\
            h_{32} &= 80 \gamma ^8-192 \gamma ^6+152 \gamma ^4-44 \gamma ^2+3\\
            h_{33} &= \gamma  \left(2 \gamma ^2-1\right) \left(64 \gamma ^6-216 \gamma ^4+258 \gamma ^2-109\right)\\
            h_{34} &= \left(2 \gamma ^2-1\right)^3 \left(5 \gamma ^2-8\right)\\
            h_{35} &= \gamma  \left(2 \gamma ^2-3\right) \left(2 \gamma ^2-1\right)^3\\
            h_{36} &= 8 \gamma ^6-28 \gamma ^4+6 \gamma ^2+3\\
            h_{37} &= \gamma  \left(384 \gamma ^8-1528 \gamma ^6+384 \gamma ^4+2292 \gamma ^2-1535\right)\\
            h_{38} &= 393897472 \gamma ^{16}-791542442 \gamma ^{14}-3429240286 \gamma ^{12}+3966858415 \gamma ^{10}\\
            &\quad+767410066 \gamma ^8-21241500 \gamma ^6+7188300 \gamma ^4-1837500 \gamma ^2+385875\\
            h_{39} &= 1575 \gamma ^7-2700 \gamma ^6-3195 \gamma ^5+3780 \gamma ^4+4993 \gamma ^3-1188 \gamma ^2-1485 \gamma +108\\
            h_{40} &= -3592192 \gamma ^{18}+2662204 \gamma ^{16}+46406238 \gamma ^{14}-37185456 \gamma ^{12}-25426269 \gamma ^{10}\\
            &\quad+222810 \gamma ^8-246540 \gamma ^6+79800 \gamma ^4-19950 \gamma ^2+3675\\
            h_{41} &= 44 \gamma ^6-32 \gamma ^4-425 \gamma ^2-82\\
            h_{42} &= \gamma  \left(16 \gamma ^6+24 \gamma ^4-226 \gamma ^2-151\right)\\
            h_{43} &= \gamma ^2 \left(4 \gamma ^8-59 \gamma ^4+35 \gamma ^2+60\right)\\
            h_{44} &= -525 \gamma ^7+1065 \gamma ^5-3883 \gamma ^3+1263 \gamma\\
            h_{45} &= 175 \gamma ^7-150 \gamma ^6-355 \gamma ^5+210 \gamma ^4+185 \gamma ^3-66 \gamma ^2-37 \gamma +6\\
            h_{46} &= -175 \gamma ^7+355 \gamma ^5-185 \gamma ^3+37 \gamma\\
            h_{47} &= \gamma  \left(525 \gamma ^6-1065 \gamma ^4-2773 \gamma ^2+1041\right)\\
            h_{48} &= 96 \gamma ^{10}-8464 \gamma ^8+54616 \gamma ^6-70104 \gamma ^4+9916 \gamma ^2+13895\\
            h_{49} &= 6144 \gamma ^{16}-587336 \gamma ^{14}+4034092 \gamma ^{12}-417302 \gamma ^{10}-5560073 \gamma ^8-142640 \gamma ^6\\
            &\quad+35710 \gamma ^4-8250 \gamma ^2+1575\\
            h_{50} &= -3747 \gamma ^6+3249 \gamma ^4+8535 \gamma ^2+1051\\
            h_{51} &= 24576 \gamma ^{18}+213480 \gamma ^{16}-1029342 \gamma ^{14}-1978290 \gamma ^{12}+3752006 \gamma ^{10}+816595 \gamma ^8\\
            &\quad-55260 \gamma ^6+13690 \gamma ^4-3100 \gamma ^2+525\\
            h_{52} &= \gamma  \left(16 \gamma ^6+204 \gamma ^4-496 \gamma ^2-869\right)\\
            h_{53} &= \gamma ^2 \left(8 \gamma ^4-6 \gamma ^2-9\right)\\
            h_{54} &= \gamma  \left(2 \gamma ^2-3\right) \left(8 \gamma ^6-6 \gamma ^4-51 \gamma ^2-8\right)\\
            h_{55} &= -4321 \gamma ^6+3387 \gamma ^4+15261 \gamma ^2+2057\\
            h_{56} &= 2100 \gamma ^7-4996 \gamma ^6+1755 \gamma ^5+4332 \gamma ^4-6422 \gamma ^3+4212 \gamma ^2-1209 \gamma +36\\
            h_{57} &= -1249 \gamma ^6+1083 \gamma ^4+1053 \gamma ^2+9\\
            h_{58} &= -1823 \gamma ^6+1221 \gamma ^4+13155 \gamma ^2+2039\\
            h_{59} &= -24 \gamma ^6+18 \gamma ^4+111 \gamma ^2+16\\
            h_{60} &= \gamma  \left(26 \gamma ^2-9\right)
          \end{aligned}\nn
        \end{equation}\\
        \hline
      
    \end{tabular}
    }
    \caption{Polynomials entering  the total impulse as well as conservative/dissipative parts (see the supplemental material).}
    \label{tbl:coeffs}
    \vspace{-0.7cm}
  \end{table*}
\end{widetext}

 {\bf Radiated energy/momentum.}
From the impulse we derive the change in the mechanical momentum of the system (in the incoming center-of-mass), which gives us the total radiated momentum, $P^\mu_{\rm rad}=-(\Delta p^\mu_1+\Delta p_2^\mu)$.  The radiated energy for hyperbolic motion at 4PM, given by $\Delta E_{\rm hyp} \equiv P_{\rm rad}\cdot \frac{m_1 u_1 + m_2 u_2 }{M \Gamma}$ with $\Gamma\equiv \!\tfrac{E}{M}$ ($M, E$ the total mass and energy and $\nu \equiv \tfrac{m_1m_2}{M^2}$), then becomes:
\begin{widetext}
  \begin{equation}
    \resizebox{0.95\hsize}{!}{
      $
      \begin{aligned}
        \label{de4}
        \Delta E_{\rm hyp}^{\rm 4PM} 
        &= \begin{multlined}[t]
          -\frac{G^4 M^5 \nu ^2}{b^4 \Gamma }\Bigg\{
          \frac{15 \pi ^2 \left(\gamma ^2-1\right) \left(27 \left(\gamma ^2-1\right) h_{31}+2 h_{50}\right)+64 \left(45 h_{32}-h_{48}\right)}{1440 \left(\gamma ^2-1\right)^3}+\frac{h_{49}}{1440 \gamma ^7 \left(\gamma ^2-1\right)^{5/2}}
          -\arccosh^2(\gamma ) \left(\frac{16 h_{53}}{\left(\gamma ^2-1\right)^2}+\frac{32 h_{54}}{\left(\gamma ^2-1\right)^{7/2}}\right)\\
          -\frac{h_{55} \log (2)\arccosh(\gamma )}{4 \left(\gamma ^2-1\right)^2}
          +\frac{h_{57} \log \left(\frac{2}{\gamma +1}\right)\arccosh(\gamma )}{4 \left(\gamma ^2-1\right)^2}
          -\frac{h_{58} \log (\gamma )\arccosh(\gamma )}{4 \left(\gamma ^2-1\right)^2}
          +\arccosh(\gamma ) \left(\frac{h_{51}}{480 \gamma ^8 \left(\gamma ^2-1\right)^3}-\frac{16 h_{52}}{5 \left(\gamma ^2-1\right)^{3/2}}\right)\\
          -\frac{h_{56} \text{Li}_2\left(\sqrt{\frac{\gamma -1}{\gamma +1}}\right)}{8 \left(\gamma ^2-1\right)^2}
          +\frac{h_{56} \text{Li}_2\left(\frac{\gamma -1}{\gamma +1}\right)}{32 \left(\gamma ^2-1\right)^2}+\frac{h_{57} \text{Li}_2\left(\sqrt{\gamma ^2-1}-\gamma \right)}{2 \left(\gamma ^2-1\right)^2}
          +\frac{h_{58} \text{Li}_2\left(-\left(\gamma -\sqrt{\gamma ^2-1}\right)^2\right)}{8 \left(\gamma ^2-1\right)^2}\\
          +\nu\Bigg[
            \frac{4 \left(-45 h_{32}+30 h_{33}-30 h_{37}+h_{48}\right)}{45 \left(\gamma ^2-1\right)^3}+\frac{\pi ^2 \left(54 \left(\gamma ^2-1\right) h_{31}+h_{39}-4 h_{50}\right)}{96 \left(\gamma ^2-1\right)^2}
            -\arccosh^2(\gamma ) \left(\frac{16 \left(h_{42}-2 h_{53}\right)}{\left(\gamma ^2-1\right)^2}-\frac{64 (h_{43}+h_{54})}{\left(\gamma ^2-1\right)^{7/2}}\right)\\+\frac{h_{38}-490 \gamma  \left(3840 \gamma ^7 h_{34}+h_{49}\right)}{352800 \gamma ^8 \left(\gamma ^2-1\right)^{5/2}}
           +\frac{(3 h_{46}+2h_{57}) \log \left(\frac{\gamma +1}{2}\right)\arccosh(\gamma )}{4 \left(\gamma ^2-1\right)^2}
            +\frac{\left(h_{44}+2 h_{55}\right) \log (2)\arccosh(\gamma )}{4 \left(\gamma ^2-1\right)^2}
            +\frac{\left(h_{47}+2 h_{58}\right) \log (\gamma )\arccosh(\gamma )}{4 \left(\gamma ^2-1\right)^2}\\
            +\arccosh(\gamma ) \left(\frac{53760 \gamma ^9 h_{35}-14 \gamma  h_{51}+h_{40}}{3360 \gamma ^9 \left(\gamma ^2-1\right)^3}-\frac{32 \left(15 h_{36}-10 h_{41}-3 h_{52}\right)}{15 \left(\gamma ^2-1\right)^{3/2}}\right)
            +\frac{\left(h_{56}-6 h_{45}\right) \text{Li}_2\left(\sqrt{\frac{\gamma -1}{\gamma +1}}\right)}{4 \left(\gamma ^2-1\right)^2}-\frac{\left(h_{56}-6 h_{45}\right) \text{Li}_2\left(\frac{\gamma -1}{\gamma +1}\right)}{16 \left(\gamma ^2-1\right)^2}\\-\frac{\left(3 h_{46}+2 h_{57}\right) \text{Li}_2\left(\sqrt{\gamma ^2-1}-\gamma \right)}{2 \left(\gamma ^2-1\right)^2}
            -\frac{\left(h_{47}+2 h_{58}\right) \text{Li}_2\left(-\left(\gamma -\sqrt{\gamma ^2-1}\right)^2\right)}{8 \left(\gamma ^2-1\right)^2}
            \Bigg]
          \Bigg\}\,.
        \end{multlined}
      \end{aligned}
      $
    }
  \end{equation}
\end{widetext}
Expanding in small velocities ($v_\infty^2 = \gamma^2-1 \ll 1$) we find, 
\beq
\label{DE}
\resizebox{0.88\hsize}{!}{
      $\begin{aligned}
 &\frac{b^4 \Delta E_{\rm hyp}^{\rm 4PM}}{G^4M^5\nu^2} = \frac{1568}{45 v_{\infty }}+\left(\frac{18608}{525}-\frac{1136 \nu }{45}\right) v_{\infty }+\frac{3136 v_{\infty }^2}{45}\\ &+ \left(\frac{764 \nu ^2}{45}-\frac{356 \nu }{63}+\frac{220348}{11025}\right) v_{\infty }^3+\left(\frac{1216}{105}-\frac{2272 \nu }{45}\right) v_{\infty }^4 \\ &+ \left(-\frac{622 \nu  ^3}{45}+\frac{3028 \nu ^2}{1575}-\frac{199538 \nu }{33075}-\frac{151854}{13475}\right) v_{\infty }^5 \\ &+\left(\frac{1528 \nu ^2}{45}-\frac{8056 \nu }{1575}+\frac{117248}{1575}\right) v_{\infty }^6+O\left(v_{\infty }^7\right)\,.
   \end{aligned}
      $
    }
    \eeq
A notable feature of the 4PM momentum is the emergence of a total recoil along the perpendicular direction. The full result is rather lengthy, however, performing a velocity expansion we find (with $\Delta_m \equiv \tfrac{m_1-m_2}{M}$)   
\beq
\resizebox{0.88\hsize}{!}{
      $\begin{aligned}
      \label{pb}
\frac{b^4 P^{\rm 4PM}_{b,\rm rad}}{\pi \Delta_m  G^4M^5 \nu^2}&=\frac{37  }{30 }+\frac{1661   v_{\infty }^2}{560 }+\frac{1491   v_{\infty }^3}{400 }
+\frac{23563 v_{\infty }^4}{10080 } \\  &-\frac{26757   v_{\infty }^5}{5600 }+\frac{700793   v_{\infty }^6}{506880 } +O(v_{\infty }^7)\,.
   \end{aligned}
      $
    }
    \eeq 
 Both expressions in \eqref{DE} and \eqref{pb} are consistent with the state-of-the-art in the PN expansion \cite{Cho:1,Cho:2,Bini:2022yrk,Bini:2021qvf,Bini:2022xpp,privateDD}. On the other hand, in the opposite limit, as $\gamma$ gets large, 
\bea
\label{divE}
\frac{b^4 \Gamma \Delta E_{\rm hyp}^{\rm 4PM}}{G^4M^5\nu^2} &\to&\frac{13696}{105} \gamma ^3 \nu \log (2 \gamma)\,,
\eea
which signals the presence of (logarithmic) mass singularities. We return to these limits in the conclusions. 
\vskip 4pt {\bf GW energy flux.} The B2B map allows us to connect the radiated energy for the scattering process to its counterpart over a period of an elliptic-like orbit via $\Delta E_{\rm ell}(j) = \Delta E_{\rm hyp}(j) - \Delta E_{\rm hyp}(-j)$~\cite{b2b3},
where $j \equiv \tfrac{p_\infty b}{GM^2\nu}$, with $p_\infty \equiv \tfrac{M \nu}{\Gamma}v_\infty$, is the (reduced) angular momentum. However, similarly to the periastron advance at ${\cal O}(G^3)$ \cite{paper1,paper2}, this expression vanishes at 4PM. To obtain radiative observables for generic orbits we derived
instead the energy flux.  Since nonlinear radiation-reaction terms do not contribute to the energy loss at this order we can resort to an adiabatic expansion. By writing the PM-expanded energy flux in an isotropic gauge as \cite{b2b3}
\beq
      \begin{aligned}
   \frac{dE}{dt} =  \frac{M}{r}\sum_n  {\cal F}^{(n)}_E(\gamma) \left(\frac{GM}{r}\right)^{(n+3)}\,,
    \end{aligned}
    \eeq
we find at 4PM order (see \cite{b2b3} for the 3PM term)
\beq
\resizebox{0.88\hsize}{!}{
      $\begin{aligned}
M\pi\xi \, {\cal F}_E^{(1)} &= \frac{3 \pi  \Gamma ^2 \nu}{4 \left(\gamma ^2-1\right)^{3/2}}\, \Delta E_{j\,\textrm{hyp}}^{(1)} 
-\frac{2 \nu^3\Delta E_{j\, \textrm{hyp}}^{(0)}}{\left(\gamma ^2-1\right)^2 \Gamma ^6 \xi ^2}\\ & \times\big[(\gamma -1)^3 \left(10 \gamma ^3-10 \gamma ^2-9 \gamma +5\right) \nu ^2\\
    &+4 \left(5 \gamma ^5-8 \gamma ^4+\gamma ^3+4 \gamma ^2-3 \gamma +1\right) \nu\\ &+8 \gamma ^4-4 \gamma ^2-1\big]\,,
   \end{aligned}
      $
    }
    \eeq
with  $\Delta { E}(j) \!= \!\sum_{n=0}^{\infty} \frac{\Delta E_j^{(n)}}{j^{n+3}}$,  $\xi  \!\equiv  \! \tfrac{E_1E_2}{E^2}\,,   E_a  \!\equiv \sqrt{p_\infty^2+m_a^2}$. \vskip 8pt 

{\bf Conclusions.}  We completed the knowledge of the total relativistic impulse in the scattering of non-spinning bodies at 4PM order, including linear, nonlinear, and hereditary radiation-reaction effects. We also derived~the total radiated spacetime momentum at ${\cal O}(G^4)$, from which we extracted the energy flux. In combination with conservative results in \cite{4pmeft,4pmeft2}, the GW flux can~be used to compute observables for generic (un)bound orbits, incorporating an infinite series of velocity corrections. The most intricate part of the calculation involves a series of master integrals with retarded propagators, which we are able to compute to all orders in the velocity through the methodology of differential equations, without resorting to PN resummations. The~boundary conditions in the near-static limit are obtained via the method of regions, thus making direct contact with derivations in the PN regime with potential and radiation modes. We find perfect agreement with various partial calculations in the literature. Explicit values can be found in the supplemental material and ancillary~file.\vskip 4pt 

There are still some key aspects of the structure of the impulse at 4PM order which deserve further study. Firstly, concerning the high-energy limit, although many nontrivial cancelations occur we find that the impulse does not transition smoothly, but rather it diverges when $m_a \to 0$ while $\gamma m_1 m_2$ is held fixed. Yet, we find that all integrals of the form in \eqref{4pm-ints} vanish in dimensional regularization for the case of incoming velocities obeying the null condition ($u_a^2=0$). This implies that the divergent terms in the $m_a\to 0$ limit are associated with the enforcement from the onset of timelike worldlines ($u_a^2 = 1$) in the massive theory. Moreover, similarly to what happens at 3PM, the radiated energy diverges, this time with a factor of $\log(2\gamma)$ in the $\gamma \gg 1$ limit at 4PM order. We~expect this behavior to be tamed in the nonperturbative solution (see e.g. \cite{Gruzinov:2014moa,DiVecchia:2022nna} and references therein).\vskip 4pt Secondly, there is the issue of the mass scaling of the impulse \cite{Bini:2021gat}, and violations thereof, e.g.~\cite{Blumlein:2021txe,Almeida:2022jrv}. As~we mentioned, some of the radiative contributions affect only the total radiated momentum in the $\hat b$-direction, while conserving energy. Moreover, they are even under time reversal, see \eqref{pb}. To~gain intuition about these terms, from the impulse and total recoil we derived the relative deflection angle (see supplemental material).  We~find perfect consistency with the (odd-in-velocity) PN values in \cite{Bini:2021gat}. Yet, starting at 5PN order, we encounter conservative-like (even-in-velocity) ${\cal O}(\nu^2)$ contributions beyond the Feynman-only part.\footnote{The deflection angle is however in tension with the two distinct values reported in \cite{Blumlein:2021txe,Almeida:2022jrv}.} In principle, depending on their origin, these terms could be incorporated into a {\it relative} Hamiltonian. We~will discuss these issues in more detail elsewhere.
 
\vskip 4pt In summary, in addition to the natural connections to GW science, e.g.~\cite{Khalil:2022ylj, Hopper:2022rwo}, the solution of the (classical) relativistic scattering problem at ${\cal O}(G^4)$ presented here demonstrates how the worldline EFT approach \cite{review,Goldberger:2022ebt}, combined~with the methodology of differential equations and integration by regions---already successfully implemented both in the conservative  \cite{pmeft,3pmeft,tidaleft,pmefts,4pmeft,4pmeft2} and dissipative \cite{Mougiakakos:2021ckm,Riva:2021vnj,Mougiakakos:2022sic,Riva:2022fru,eftrad} sectors---are very powerful tools to tackle the entire two-body dynamics in general relativity within the PM expansion. Complete results at higher PM orders, including spin and tidal effects, are underway.\vskip 8pt

 {\bf Acknowledgements.} We thank Donato Bini, Johannes Bl\"umlein, Ruth Britto, Gihyuk Cho, Walter Goldberger, Carlo Heissenberg, Francois Larrouturou, Massimiliano Riva, Ira Rothstein, Rodolfo Russo, and Chia-Hsien Shen for useful discussions. We are particularly grateful to Donato Bini, Thibault Damour and Andrea Geralico for sharing their results on higher order radiative effects in the PN expansion prior to submission. We would like to thank also the participants of the program ``High-Precision Gravitational Waves" at the KITP (supported in part by the National Science Foundation under Grant No. NSF PHY-1748958) for discussions. ZL is grateful to the Mainz Institute for Theoretical Physics of the DFG Cluster of Excellence PRISMA$^+$ (Project ID 39083149), for its hospitality and support. This~work received support from the ERC-CoG  {\it Precision Gravity: From the LHC to LISA} provided by the European Research Council (ERC) under the European Union's H2020 research and innovation programme (grant No.\,817791). 	 ZL is supported partially by DFF grant 1026-00077B, the Carlsberg foundation, and the European Union’s Horizon 2020 research and innovation program under the Marie Sk\l{}odowska-Curie grant agreement No.\,847523 `INTERACTIONS’.

\newpage
\begin{widetext} 
\section{Supplemental Material}
We give explicit values for the conservative and dissipative contributions to the impulse as well as the (relative) scattering angle. Ready-to-use results can also be found in the ancillary file.\vskip 4pt  {\it Impulse}. For the conservative part we find
\beq
\label{impcon}
\resizebox{0.94\hsize}{!}{
  $
  \begin{aligned}
    \frac{c^{(4)\rm cons}_{1b}}{\pi} &= \begin{multlined}[t]
      -\frac{3 h_1 m_1 m_2 \left(m_1^3+m_2^3\right)}{64 \left(\gamma ^2-1\right)^{5/2}}
      +m_1^2 m_2^2 \left(m_1+m_2\right)\bigg[
        \frac{21 \sqrt{\gamma ^2-1} h_2 \mathrm{E}^2\left(\frac{\gamma -1}{\gamma +1}\right)}{32 (\gamma -1)^2 (\gamma +1)}
        +\frac{3 h_3 \mathrm{K}^2\left(\frac{\gamma -1}{\gamma +1}\right)}{16 \left(\gamma ^2-1\right)^{3/2}}
        -\frac{3 h_4 \mathrm{E}\left(\frac{\gamma -1}{\gamma +1}\right) \mathrm{K}\left(\frac{\gamma -1}{\gamma +1}\right)}{16 \left(\gamma ^2-1\right)^{3/2}}
        +\frac{\pi ^2  h_5}{8 \sqrt{\gamma ^2-1}}\\
        +\frac{h_6 \log \left(\frac{\gamma -1}{2}\right)}{16 \left(\gamma ^2-1\right)^{3/2}}
        +\frac{3 h_7 \text{Li}_2\left(\sqrt{\frac{\gamma -1}{\gamma +1}}\right)}{(\gamma -1) (\gamma +1)^2}
        -\frac{3 h_7 \text{Li}_2\left(\frac{\gamma -1}{\gamma +1}\right)}{4 (\gamma -1) (\gamma +1)^2}
        -\frac{3 h_{15} \log \left(\frac{\gamma -1}{2}\right) \log \left(\frac{\gamma +1}{2}\right)}{8 \sqrt{\gamma ^2-1}}
        +\frac{3 h_{16} \log \left(\frac{\gamma -1}{2}\right) \arccosh(\gamma )}{16 \left(\gamma ^2-1\right)^2}
        +\frac{h_{20}}{192 \gamma ^7 \left(\gamma ^2-1\right)^{5/2}}\\
        +\frac{h_{22} \log \left(\frac{\gamma +1}{2}\right)}{16 \left(\gamma ^2-1\right)^{3/2}}
        +\frac{h_{23} \log (\gamma )}{2 \left(\gamma ^2-1\right)^{3/2}}
        -\frac{h_{24} \arccosh(\gamma )}{16 \left(\gamma ^2-1\right)^3}
        -\frac{3 h_{26} \arccosh^2(\gamma )}{32 \left(\gamma ^2-1\right)^{7/2}}
        +\frac{3 h_{27} \log ^2\left(\frac{\gamma +1}{2}\right)}{2 \sqrt{\gamma ^2-1}}
        +\frac{3 h_{28} \log \left(\frac{\gamma +1}{2}\right) \arccosh(\gamma )}{16 \left(\gamma ^2-1\right)^2}
        +\frac{h_{29} \text{Li}_2\left(\frac{1-\gamma }{\gamma +1}\right)}{4 \sqrt{\gamma ^2-1}}\\
        +\frac{3 \sqrt{\gamma ^2-1} h_{30} \text{Li}_2\left(\frac{\gamma -1}{\gamma +1}\right)}{8 (\gamma -1) (\gamma +1)}
        \bigg]
    \end{multlined}\\
    c^{(4)\rm cons}_{1\check{u}_1} &=
    \frac{9 \pi ^2 h_{31} m_1 m_2^2 \left(m_1+m_2\right){}^2}{32 \left(\gamma ^2-1\right)}
    +\frac{2 h_{32} m_1 m_2^2 \left(m_1^2+m_2^2\right)}{\left(\gamma ^2-1\right)^3}
    +m_1^2 m_2^3\bigg[
      \frac{4 h_{33}}{3 \left(\gamma ^2-1\right)^3}
      -\frac{16 h_{36} \arccosh(\gamma )}{\left(\gamma ^2-1\right)^{3/2}}
      \bigg] \\
    c^{(4)\rm cons}_{1\check{u}_2} &=
    -\frac{9 \pi ^2 h_{31} m_1^2 m_2 \left(m_1+m_2\right){}^2}{32 \left(\gamma ^2-1\right)}
    -\frac{2 h_{32} m_1^2 m_2 \left(m_1^2+m_2^2\right)}{\left(\gamma ^2-1\right)^3}
    +m_1^3 m_2^2\bigg[
      -\frac{4 h_{33}}{3 \left(\gamma ^2-1\right)^3}
      +\frac{16 h_{36} \arccosh(\gamma )}{\left(\gamma ^2-1\right)^{3/2}}
    \bigg]
\end{aligned}
$
}
\eeq
which reproduces the result in \cite{4pmeft,4pmeft2}.  For the dissipative part we split the result into two terms. For the (instantaneous) contribution involving a single radiation mode we have
\beq
\label{impdis1}
\resizebox{0.94\hsize}{!}{
  $
  \begin{aligned}
    \frac{c^{(4) \rm diss}_{1b,\rm 1rad}}{\pi} &= \begin{multlined}[t]
      m_1^3 m_2^2\bigg[
        \frac{h_8}{48 \left(\gamma ^2-1\right)^3}
        +\frac{h_{10} \log \left(\frac{\gamma +1}{2}\right)}{8 \left(\gamma ^2-1\right)^2}
        +\frac{h_{14} \arccosh(\gamma )}{16 \left(\gamma ^2-1\right)^{7/2}}
        \bigg]
      +m_1^2 m_2^3\bigg[
        \frac{h_{19}}{48 \left(\gamma ^2-1\right)^3}
        +\frac{h_{21} \log \left(\frac{\gamma +1}{2}\right)}{8 \left(\gamma ^2-1\right)^2}
        +\frac{h_{25} \arccosh(\gamma )}{16 \left(\gamma ^2-1\right)^{7/2}}
        \bigg]\,,
    \end{multlined}\\
    c^{(4) \rm diss}_{1\check{u}_1,\rm 1rad} &= m_1^2 m_2^3 \bigg[
      -\frac{8 h_{34}}{3 \left(\gamma ^2-1\right)^{5/2}}
      +\frac{8 h_{35} \arccosh(\gamma )}{\left(\gamma ^2-1\right)^3}
      \bigg]\,,\\
    c^{(4)\rm diss}_{1\check{u}_2,\rm 1rad} &= \begin{multlined}[t]
      m_1^3 m_2^2\bigg[
        \frac{h_{38}}{705600 \gamma ^8 \left(\gamma ^2-1\right)^{5/2}}
        +\frac{\pi ^2 \left(108 h_{31}(\gamma^2-1)+h_{39}\right)}{192 \left(\gamma ^2-1\right)^2}
        +\frac{h_{40} \arccosh(\gamma )}{6720 \gamma ^9 \left(\gamma ^2-1\right)^3}
        +\frac{32 h_{43} \arccosh^2(\gamma )}{\left(\gamma ^2-1\right)^{7/2}}
        +\frac{h_{44} \log (2) \arccosh(\gamma )}{8 \left(\gamma ^2-1\right)^2}\\
        +\frac{3 h_{45} \left(\text{Li}_2\left(\frac{\gamma -1}{\gamma +1}\right)-4 \text{Li}_2\left(\sqrt{\frac{\gamma -1}{\gamma +1}}\right)\right)}{16 \left(\gamma ^2-1\right)^2}
        +\frac{3 h_{46} \left(\log \left(\frac{\gamma +1}{2}\right) \arccosh(\gamma )-2 \text{Li}_2\left(\sqrt{\gamma ^2-1}-\gamma \right)\right)}{8 \left(\gamma ^2-1\right)^2}\\
        -\frac{h_{47} \left(\text{Li}_2\left(-\left(\gamma -\sqrt{\gamma ^2-1}\right)^2\right)-2 \log (\gamma ) \arccosh(\gamma )\right)}{16 \left(\gamma ^2-1\right)^2}
        +\frac{8 h_{60} \arccosh^2(\gamma )}{\left(\gamma ^2-1\right)^2}
        \bigg]
      +m_1^2 m_2^3\bigg[
        \frac{h_{49}}{1440 \gamma ^7 \left(\gamma ^2-1\right)^{5/2}}\\
        +\frac{\pi ^2 \left(27 h_{31}(\gamma^2-1)+2 h_{50}\right)}{96 \left(\gamma ^2-1\right)^2}
        +\frac{h_{51} \arccosh(\gamma )}{480 \gamma ^8 \left(\gamma ^2-1\right)^3}
        -\frac{32 h_{54} \arccosh^2(\gamma )}{\left(\gamma ^2-1\right)^{7/2}}
        -\frac{h_{55} \log (2) \arccosh(\gamma )}{4 \left(\gamma ^2-1\right)^2}
        +\frac{h_{56} \left(\text{Li}_2\left(\frac{\gamma -1}{\gamma +1}\right)-4 \text{Li}_2\left(\sqrt{\frac{\gamma -1}{\gamma +1}}\right)\right)}{32 \left(\gamma ^2-1\right)^2}\\
        -\frac{h_{57} \left(\log \left(\frac{\gamma +1}{2}\right) \arccosh(\gamma )-2 \text{Li}_2\left(\sqrt{\gamma ^2-1}-\gamma \right)\right)}{4 \left(\gamma ^2-1\right)^2}
        +\frac{h_{58} \left(\text{Li}_2\left(-\left(\gamma -\sqrt{\gamma ^2-1}\right)^2\right)-2 \log (\gamma ) \arccosh(\gamma )\right)}{8 \left(\gamma ^2-1\right)^2}
        +\frac{16 h_{59} \arccosh^2(\gamma )}{\left(\gamma ^2-1\right)^2}
        \bigg]
    \end{multlined}
  \end{aligned}
  $
}
\eeq
whereas the contribution involving two radiation modes going on-shell is given by
\beq
\label{impdis2}
\resizebox{0.86\hsize}{!}{
  $
  \begin{aligned}
    \frac{c^{(4) \rm diss}_{1b,\rm 2rad}}{\pi} &= \begin{multlined}[t]
      m_1^3 m_2^2\bigg[
        \frac{\sqrt{\gamma ^2-1} \left(h_9-4 \gamma ^2 (\gamma +1) h_{20}\right)}{768 (\gamma -1)^3 \gamma ^9 (\gamma +1)^4}
        +\frac{\log (\gamma ) \left(h_{12}-8 \left(\gamma ^2-1\right) h_{23}\right)}{16 \left(\gamma ^2-1\right)^{5/2}}
        +\frac{\arccosh(\gamma ) \left((\gamma +1) h_{24}-2 (\gamma -1)^2 h_{13}\right)}{16 (\gamma -1)^3 (\gamma +1)^4}\\
        +\frac{3 h_{26} \arccosh^2(\gamma )}{32 \left(\gamma ^2-1\right)^{7/2}}
        +\frac{3 \left(h_{15}-4 h_{27}\right) \log ^2\left(\frac{\gamma +1}{2}\right)}{8 \sqrt{\gamma ^2-1}}
        +\log \left(\frac{\gamma +1}{2}\right) \left(\frac{-2 \left(\gamma ^2-1\right) h_{22}-h_{11}}{32 \left(\gamma ^2-1\right)^{5/2}}-\frac{3 \left(h_{16}+h_{28}\right) \arccosh(\gamma )}{16 \left(\gamma ^2-1\right)^2}\right)\\
        +\frac{\left(-3 \left(\gamma ^2-1\right) h_{18}-8 h_{29}\right) \text{Li}_2\left(\frac{1-\gamma }{\gamma +1}\right)}{32 \sqrt{\gamma ^2-1}}
        -\frac{3 \left(h_{17}+8 h_{30}\right) \text{Li}_2\left(\frac{\gamma -1}{\gamma +1}\right)}{64 \sqrt{\gamma ^2-1}}
        \bigg]\,,
  \end{multlined}\\
c^{(4) \rm diss}_{1\check{u}_1,\rm 2rad} &=0\,,\\
c^{(4)\rm diss}_{1\check{u}_2,\rm 2rad} &= \begin{multlined}[t]
  m_1^3 m_2^2\bigg[
    \frac{4 \left(h_{33}-h_{37}\right)}{3 \left(\gamma ^2-1\right)^3}
    -\frac{16 \left(3 h_{36}-2 h_{41}\right) \arccosh(\gamma )}{3 \left(\gamma ^2-1\right)^{3/2}}
    -\frac{8 \left(h_{42}+h_{60}\right) \arccosh^2(\gamma )}{\left(\gamma ^2-1\right)^2}
    \bigg]\\
  +m_1^2 m_2^3\bigg[
    \frac{2 \left(45 h_{32}-h_{48}\right)}{45 \left(\gamma ^2-1\right)^3}
    -\frac{16 h_{52} \arccosh(\gamma )}{5 \left(\gamma ^2-1\right)^{3/2}}
    -\frac{16 \left(h_{53}+h_{59}\right) \arccosh^2(\gamma )}{\left(\gamma ^2-1\right)^2}
    \bigg]
  \end{multlined}
\end{aligned}
$
}
\eeq
We find perfect agreement with various partial results reported in \cite{Damour:2020tta,Cho:1,Cho:2,Bini:2021gat,Bini:2021qvf,Bini:2022yrk,Bini:2022xpp,privateDD}.\vskip 4pt

{\it Scattering angle.} By performing a Lorentz boost using the value of the total recoil we can derive the relative spacelike impulse, $\Delta \bp$, via (see  \cite{Damour:2020tta}) 
\beq
\label{relative}
\Delta \bp =  \Delta \bp_1 + \frac{E_1}{E} \bP_{\rm rad} + {\cal O}\big(\bP_{\rm rad}^2\big)\,.
\eeq
From here we can obtain a relative scattering angle, given by (recall $j \equiv p_\infty b/(GM^2\nu)$)
 \beq \frac{\chi_{\rm rel}}{2} \equiv \frac{1}{2} \mathrm{arccos}\left(\frac{\bp_+\cdot \bp_- }{ |{\bp}_-||\bp_+|}\right) = \sum_{n=1}^{\infty} \chi^{(n)}_{b,\rm rel} \left(\frac{GM}{b}\right)^n  = \sum_{n=1}^{\infty} \frac{\chi^{(n)}_{j,\rm rel} }{j^n}  \,, \eeq with $\bp_-$ and $\bp_+ \equiv \bp_- + \Delta \bp$, the relative incoming/outgoing three-momentum at infinity.\vskip 4pt We find, in impact parameter space,
\begin{align}
\label{anglerel}
\chi^{(4)}_{b, \rm rel}(\gamma) &=  \chi^{(4)\rm cons}_{b,\rm rel}(\gamma) + \delta  \chi^{(4)\rm rr}_{b,\rm rel}(\gamma)\\
\delta  \chi^{(4)\rm rr}_{b,\rm rel}(\gamma) &= \chi^{(4)\rm 1rad}_{b,\rm rel}(\gamma) +  \chi^{(4)\rm 2rad}_{b,\rm rel}(\gamma)\,,
\end{align}
where 
\beq
\label{anglecons}
\resizebox{0.94\hsize}{!}{
  $
  \frac{ \chi^{(4)\rm cons}_{b,\rm rel}(\gamma)}{\pi\Gamma}= \begin{multlined}[t]
    \frac{3 h_{61}}{128 \left(\gamma ^2-1\right)^3}
    +\nu\bigg[
      -\frac{3 h_3 \mathrm{K}^2\left(\frac{\gamma -1}{\gamma +1}\right)}{32 \left(\gamma ^2-1\right)^2}
      +\frac{3 h_4 \mathrm{E}\left(\frac{\gamma -1}{\gamma +1}\right) \mathrm{K}\left(\frac{\gamma -1}{\gamma +1}\right)}{32 \left(\gamma ^2-1\right)^2}
      +\frac{\pi ^2 h_5}{16(1- \gamma ^2)}
      +\frac{3 h_{27} \log ^2\left(\frac{\gamma +1}{2}\right)}{4(1- \gamma ^2)}
      -\frac{h_6 \log \left(\frac{\gamma -1}{2}\right)}{32 \left(\gamma ^2-1\right)^2}
      +\frac{3 h_{15} \log \left(\frac{\gamma -1}{2}\right) \log \left(\frac{\gamma +1}{2}\right)}{16 \left(\gamma ^2-1\right)}\\
      -\frac{h_{22} \log \left(\frac{\gamma +1}{2}\right)}{32 \left(\gamma ^2-1\right)^2}
      -\frac{h_{23} \log (\gamma )}{4 \left(\gamma ^2-1\right)^2}
      +\frac{3 h_{26} \arccosh^2(\gamma )}{64 \left(\gamma ^2-1\right)^4}
      +\frac{h_{24} \arccosh(\gamma )}{32 \left(\gamma ^2-1\right)^{7/2}}
      -\frac{3 h_{16} \log \left(\frac{\gamma -1}{2}\right) \arccosh(\gamma )}{32 \left(\gamma ^2-1\right)^{5/2}}
      -\frac{3 h_{28} \log \left(\frac{\gamma +1}{2}\right) \arccosh(\gamma )}{32 \left(\gamma ^2-1\right)^{5/2}}\\
      -\frac{h_{62}}{384 \gamma ^7 \left(\gamma ^2-1\right)^3}
      -\frac{21 h_2 \mathrm{E}^2\left(\frac{\gamma -1}{\gamma +1}\right)}{64 (\gamma -1)^2 (\gamma +1)}
      -\frac{3 \sqrt{\gamma ^2-1} h_7 \text{Li}_2\left(\sqrt{\frac{\gamma -1}{\gamma +1}}\right)}{2 (\gamma -1)^2 (\gamma +1)^3}
      +\frac{h_{29} \text{Li}_2\left(\frac{1-\gamma }{\gamma +1}\right)}{8(1- \gamma^2)}+\left(\frac{3 \sqrt{\gamma ^2-1} h_7}{8 (\gamma -1)^2 (\gamma +1)^3}
      +\frac{3 h_{30}}{16-16 \gamma ^2}\right) \text{Li}_2\left(\frac{\gamma -1}{\gamma +1}\right)
      \bigg]\,,
    \end{multlined}
  $
  }
\eeq
which agrees with the Feynman-only part computed in \cite{4pmeft,4pmeft2} (including potential and tail terms), and  
\beq
\label{anglediss}
\resizebox{0.85\hsize}{!}{
  $
  \begin{aligned}
  \frac{\Gamma\chi^{(4)\rm 1rad}_{b,\rm rel}(\gamma)}{\pi\nu} &= \begin{multlined}[t]
    \frac{h_{64}}{96 \left(\gamma ^2-1\right)^{7/2}}+\frac{h_{65} \log \left(\frac{\gamma +1}{2}\right)}{16 \left(\gamma ^2-1\right)^{5/2}}+\frac{h_{63} \arcsinh\left(\frac{\sqrt{\gamma -1}}{\sqrt{2}}\right)}{8 \left(\gamma ^2-1\right)^4}-\frac{h_{25} \arccosh(\gamma )}{32 \left(\gamma ^2-1\right)^4}\\
    +\nu\bigg[
      \frac{h_{67}}{96 \left(\gamma ^2-1\right)^{7/2}}+\frac{h_{68} \log \left(\frac{\gamma +1}{2}\right)}{16 \left(\gamma ^2-1\right)^{5/2}}-\frac{\arccosh(\gamma ) \left((\gamma +1) h_{14}+(\gamma -3) h_{25}\right)}{32 \left(\gamma ^2-1\right)^4}+\frac{h_{66} \arcsinh \left(\frac{\sqrt{\gamma -1}}{\sqrt{2}}\right)}{8 (\gamma -1)^2 (\gamma +1)^4}
      \bigg]\,,
    \end{multlined}\\
  \frac{\Gamma\chi^{(4)\rm 2rad}_{b,\rm rel}(\gamma)}{\pi\nu^2} &= \begin{multlined}[t]
    \frac{\log \left(\frac{\gamma +1}{2}\right) \left(2 \left(\gamma ^2-1\right) h_{22}+h_{11}\right)}{64 (\gamma -1)^3 (\gamma +1)^2}-\frac{\log (\gamma ) \left(h_{12}-8 \left(\gamma ^2-1\right) h_{23}\right)}{32 (\gamma -1)^3 (\gamma +1)^2}+\frac{\arccosh(\gamma ) \left(2 (\gamma -1)^2 h_{13}-(\gamma +1) h_{24}\right)}{32 \left(\gamma ^2-1\right)^{7/2}}\\
    +\frac{3 \sqrt{\gamma ^2-1} \left(h_{16}+h_{28}\right) \log \left(\frac{\gamma +1}{2}\right) \arccosh(\gamma )}{32 (\gamma -1)^3 (\gamma +1)^2}
    -\frac{h_9-4 \gamma ^2 (\gamma +1) h_{20}}{1536 \gamma ^9 \left(\gamma ^2-1\right)^3}
    -\frac{3 \left(h_{15}-4 h_{27}\right) \log ^2\left(\frac{\gamma +1}{2}\right)}{16 (\gamma -1)}\\
    -\frac{3 h_{26} \arccosh^2(\gamma )}{64 (\gamma -1)^4 (\gamma +1)^3}+\left(\frac{3}{64} (\gamma +1) h_{18}+\frac{h_{29}}{8 (\gamma -1)}\right) \text{Li}_2\left(\frac{1-\gamma }{\gamma +1}\right)
    +\frac{3 \left(h_{17}+8 h_{30}\right) \text{Li}_2\left(\frac{\gamma -1}{\gamma +1}\right)}{128 (\gamma -1)}\,,
    \end{multlined}
  \end{aligned}
  $
  }
\eeq
 where (for the sake of notation) we have introduced a few additional polynomials:
 \begin{equation}
   \resizebox{0.94\hsize}{!}{
     $
     \begin{aligned}
       h_{61} &= 35 (\gamma -1) (\gamma +1) \left(33 \gamma ^4-18 \gamma ^2+1\right)\\
       h_{62} &= 3600 \gamma ^{16}+4320 \gamma ^{15}-35360 \gamma ^{14}+33249 \gamma ^{13}+27952 \gamma ^{12}-25145 \gamma ^{11}-15056 \gamma ^{10}-32177 \gamma ^9+64424 \gamma ^8-38135 \gamma ^7+13349 \gamma ^6-1471 \gamma ^4\\
       &\quad+207 \gamma ^2-45\\
       h_{63} &= \gamma ^2 \left(2 \gamma ^2-3\right) \left(2 \gamma ^2-1\right) \left(35 \gamma ^4-30 \gamma ^2+11\right)\\
       h_{64} &= -4140 \gamma ^8+702 \gamma ^7+15018 \gamma ^6-8491 \gamma ^5-9366 \gamma ^4+10052 \gamma ^3-6210 \gamma ^2+2681 \gamma -102\\
       h_{65} &= 210 \gamma ^7-240 \gamma ^6-755 \gamma ^5+216 \gamma ^4+1200 \gamma ^3-508 \gamma ^2-295 \gamma +124\\
       h_{66} &= \gamma  \left(2 \gamma ^2-3\right) \left(2 \gamma ^2-1\right) \left(35 \gamma ^4-30 \gamma ^2+11\right)\\
       h_{67} &= -\left((\gamma -1) \left(420 \gamma ^9+7596 \gamma ^8-2040 \gamma ^7-30840 \gamma ^6+21667 \gamma ^5+18929 \gamma ^4-26710 \gamma ^3+14910 \gamma ^2-3177 \gamma -947\right)\right)\\
       h_{68} &= (\gamma -1) \left(490 \gamma ^7-290 \gamma ^6-1725 \gamma ^5+189 \gamma ^4+2632 \gamma ^3-952 \gamma ^2-661 \gamma +253\right)
     \end{aligned}
   $
   }
\end{equation}
\vskip 4pt
Due to the different properties under time reversal, the distinction between the two radiation-reaction terms is straightforward. This is also manifest after performing a PN expansion in small $v^2_\infty \ll 1$. In order to compare with the literature it is convenient to transform to $j$ space. We find at ${\cal O}(G^4)$:
 \begin{equation}
   \resizebox{0.85\hsize}{!}{
  $
   \begin{aligned}
\frac{\delta \chi^{(4)\rm rr}_{j,\rm rel}(\gamma)}{\pi} &= \frac{121 \nu  v_{\infty }}{30}+\left(\frac{23111 \nu }{1680}-\frac{437 \nu ^2}{60}\right) v_{\infty }^3+\left(\frac{511 \nu ^3}{48}-\frac{75253 \nu ^2}{3360}+\frac{44759 \nu }{2240}\right) v_{\infty }^5+
\frac{1491}{400} \nu ^2 v_{\infty }^6\\&+\left(-\frac{455 \nu ^4}{32}+\frac{26367 \nu ^3}{896}-\frac{288007 \nu ^2}{10080}+\frac{1350131 \nu }{1182720}\right) v_{\infty }^7+{\mathcal O}\left(v_{\infty }^8\right)\,,
  \end{aligned}
   $
   }
\end{equation}
in complete agreement with the result in Table XI of \cite{Bini:2021gat}, except for the ${\cal O}(\nu^2v_{\infty}^6)$ term. Likewise, we can PN expand the full result using the same convention as in Eq. (9.2) of \cite{Bini:2021gat}. Keeping only even-in-velocity terms, we find
 \begin{equation}
 \label{tildechi}
   \resizebox{0.9\hsize}{!}{
  $
   \begin{aligned}
\left. \frac{\Gamma^3\chi^{(4)}_{j,\rm rel}(\gamma)-  \chi^{(4)}_{j, \rm Sch}(\gamma)}{\pi}\right|_{\rm even}&=  -\frac{15 \nu }{4}+\left(\frac{123 \pi ^2 \nu }{256}-\frac{557 \nu }{16}\right) v_{\infty  }^2+ \left(\frac{33601 \pi ^2 \nu }{16384}-\frac{6113 \nu }{96}-\frac{37}{5} \nu  \log \left(v_{\infty }/2\right)\right)v_{\infty }^4\\&+ \left(\frac{1491 \nu ^2}{400}+\frac{93031 \pi ^2 \nu }{32768}-\frac{615581 \nu }{19200}-\frac{1357}{280} \nu  \log \left(v_{\infty }/2\right)\right)v_{\infty }^6 +{\mathcal O}\left(v_{\infty }^8\right)\,, \end{aligned}
   $
   }
\end{equation}
in agreement with Eq. (9.2) except for the same ${\cal O}(\nu^2 v_\infty^6)$ contribution, which first enters at 5PN order. (The~expression in \eqref{tildechi} also disagrees with the two distinct results given in \cite{Blumlein:2021txe,Almeida:2022jrv}.)\vskip 4pt  We will discuss the implications in more detail elsewhere. Let us, however, add a few remarks. 
First of all, although one could in principle incorporate these extra terms into a relative Hamiltonian, it is not entirely clear from our calculation alone which portion corresponds to nonlinear gravitational radiation-reaction forces (and/or hereditary terms), which are evaluated on the unperturbed (conservative) solution; or instead produced by effects at second order in the linear radiation-reaction force, and thus evaluated on the leading non-conservative trajectory. While we could in principle include the former the latter clearly does not belong there. Secondly, these extra terms radiate linear momentum, see \eqref{pb}. Therefore, they are only `conservative' from the point of view of the relative motion. Finally, following \cite{Bini:2021gat} we have used the relative impulse to obtain the deflection angle. Although the full analysis in \cite{Bini:2021gat} is valid at linear order in the radiation-reaction force, we have only used the Lorentz boost to the outgoing center-of-mass frame, see \eqref{relative}. In principle, this holds up to ${\cal O}(\bP_{\rm rad}^2)$, namely 6PM order. However, it is still possible that \eqref{relative} may have to be modified to fully incorporate non-linear radiation-reaction effects in the relative motion.
\end{widetext}

\end{document}